# MOLECULAR ELECTRONICS BASED ON SELF-ASSEMBLED MONOLAYERS

D. Vuillaume

## 1. INTRODUCTION

Since the first measurement of electron tunneling through an organic monolayer in 1971,(Mann and Kuhn, 1971) and the *gedanken* experiment of a molecular current rectifying diode in 1974,(Aviram and Ratner, 1974) molecular-scale electronics have attracted a growing interest, both for basic science at the nanoscale and for possible applications in nano-electronics. In the first case, molecules are quantum object by nature and their properties can be tailored by chemistry opening avenues for new experiments. In the second case, molecule-based devices are envisioned to complement silicon devices by providing new functions or already existing functions at a simpler process level and at a lower cost by virtue of their self-organization capabilities, moreover, they are not bound to von Neuman architecture and this may open the way to other architectural paradigms.

Molecular electronics, i.e. the information processing at the molecular-scale, becomes more and more investigated and envisioned as a promising candidate for the nanoelectronics of the future. One definition is "information processing using photo-, electro-, iono-, magneto-, thermo-, mechanico or chemio-active effects at the scale of structurally and functionally organized molecular architectures" (adapted from (Lehn, 1988)). In the following, we will consider devices based on organic molecules with size ranging from a single molecule to a monolayer. This definition excludes devices based on thicker organic materials referred to as organic electronics. Two works paved the foundation of this molecular-scale electronics field. In 1971, Mann and Kuhn were the first to demonstrate tunneling transport through a monolayer of aliphatic chains (Mann and Kuhn, 1971). In 1974, Aviram and Ratner theoretically proposed the concept of a molecular rectifying diode where an acceptor-bridge-donor (A-b-D) molecule can play the same role as a semiconductor p-n junction (Aviram and Ratner, 1974). Since that, many groups have reported on the electrical properties of molecular-scale devices from single molecules to monolayers.

After a brief overview of the nanofabrication of molecular devices, we review in this chapter, the electronic properties of several basic devices, from simple molecules such as molecular tunnel junctions and molecular wires, to more complex ones such as molecular rectifying diodes, molecular switches and memories.

2. NANOFABRICATION FOR MOLECULAR DEVICES

To measure the electronic transport through an organic monolayer, we need a test device as simple as possible. The generic device is a metal/monolayer/metal or metal/molecules/metal (MmM) junction (for simplicity, we will always use this term and acronym throughout the paper even if the metal electrode is replaced by a semiconductor). Organic monolayers and sub-monolayers (down to single molecules) are usually deposited on the electrodes by chemical reactions in solution or in gas phase using molecules of interest bearing a functional moiety at the ends which is chemically reactive to the considered solid surface (for instance, thiol group on metal surfaces such as Au, silane group on oxidized surfaces, etc…) – Fig. 1. However, Langmuir-Blodgett (LB) monolayers have also been used for device applications early in the 70s (see a review in a text-book (Ulman, 1991)). Some important results are, for instance, the observation of a current rectification behavior through LB monolayers of hexadecylquinolinium tricyanoquinodimethanide (Ashwell et al., 1990, Geddes et al., 1995, Martin et al., 1993, Metzger et al., 1997, Vuillaume et al., 1999, Xu et al., 2001, Metzger et al., 2001) and the fabrication of molecular switches based on LB monolayers of catenanes (Collier et al., 1999, Collier et al., 2000, Pease et al., 2001, Chen et al., 2003a, Chen et al., 2003b). The second method deals with monolayers of organic molecules chemically grafted on solid substrates, also called self-assembled monolayers (SAM) (Ulman, 1991). Many reports in the literature concern SAMs of thiol terminated molecules chemisorbed on gold surfaces, and to a less extend, molecular-scale devices based on SAMs chemisorbed on semiconductors, especially silicon. Silicon is the most widely used semiconductor in microelectronics. The capability to modify its surface properties by the chemical grafting of a broad family or organic molecules (e.g. modifying the surface potential (Bruening et al., 1994, Cohen et al., 1998, Cohen et al., 1997)) is the starting point for making almost any tailored surfaces useful for new and improved silicon-based devices. Between the end of the silicon road-map and the envisioned advent of fully molecular-scale electronics, there may be a role played by such hybrid-electronic devices (Compano et al., 2000, Joachim et al., 2000). The use of thiol-based SAMs on gold in molecular-scale electronics is supported by a wide range of experimental results on their growth, structural and electrical properties (see a review by F. Schreider (Schreiber, 2000)). However, SAMs on silicon and silicon dioxide surfaces were less studied and were more difficult to control. This has resulted in an irreproducible quality of these SAMs with large time-to-time and lab-to-lab variations. This feature may explain the smaller number of attempts to use these SAMs in molecular-scale electronics than for the thiol/gold system. Since the first chemisorption of alkyltrichlorosilane molecules from solution on a solid substrate (mainly oxidized silicon) introduced by Bigelow, Pickett and Zisman (Bigelow et al., 1946) and later developed by Maoz and Sagiv (Maoz and Sagiv, 1984), further detailed studies (Brzoska et al., 1992, Brzoska et al., 1994, Allara et al., 1995, Parikh et al., 1994) have lead to a better understanding of the basic chemical and thermodynamical mechanisms of this self-assembly process. For a review on these processes, see Refs. (Ulman, 1991, Schreiber, 2000).

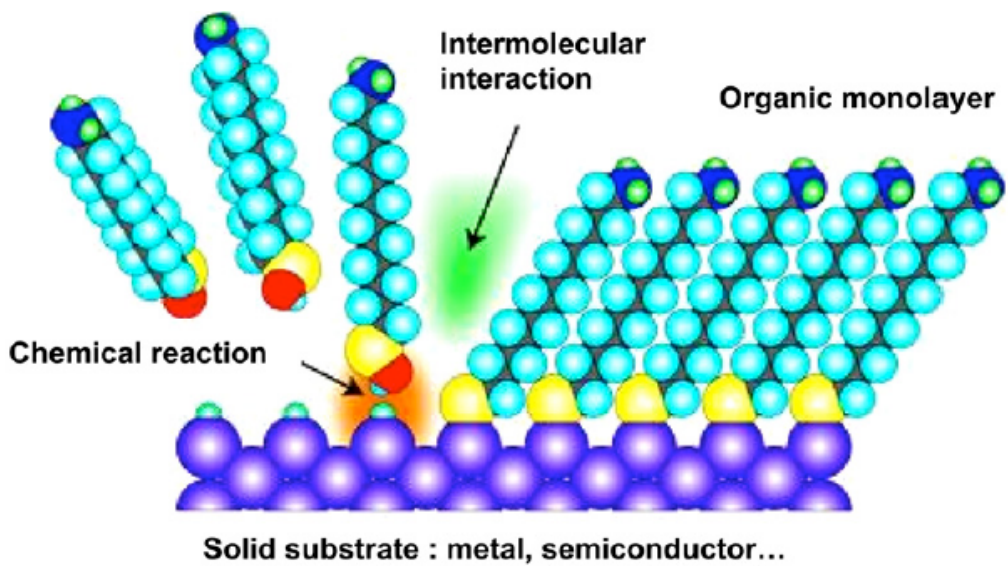

Fig. 1: A schematic description of the formation of an organic monolayer on a solid substrate, showing the chemical reaction between a functionalized end of the molecule and the substrate, and the interactions between adjacent molecules (from www.mtl.kyoto-u.ac.jp/groups/sugimura-g/index-E.html, slightly modified).

In their pioneering work, Mann and Kuhn used a mercury drop to contact the monolayer (Mann and Kuhn, 1971), and this technique is still used nowadays (Holmlin et al., 2001, Rampi et al., 1998, Selzer et al., 2002a, Selzer et al., 2002b) at the laboratory level as an easy technique for a quick assessment of the electrical properties. Several types of MmM junctions have been built. The simplest structure consists of depositing the monolayer onto the bottom electrode and then evaporating a metal electrode on top of the monolayer through a masking technique. These shadow masks are fabricated from metal or silicon nitride membranes and the dimensions of the holes in the mask may range from few hundreds of µm to few tens of nanometers. Chen and coworkers (Chen et al., 1999, Chen et al., 2000) have used nanopores (about 30 nm in diameter in a silicon nitride membrane), in which a small numbers of molecules are chemisorbed to fabricate these MmM junctions. From ~$10^{10}$ to ~$10^2$ molecules can be measured in parallel with these devices. The critical point deals with the difficult problem of making a reliable metal contact on top of an organic monolayer. Several studies (Herdt and Czanderna, 1995, Jung and Czanderna, 1994, Jung et al., 1996, Fisher et al., 2000, Fisher et al., 2002, Konstadinidis et al., 1995) have analyzed (by X-ray photoelectron spectroscopy, infra-red spectroscopy,…) the interaction (bond insertion, complexation…) between the evaporated atoms and the organic molecules in the SAM. When the metal atoms are strongly reactive with the end-groups of the molecules (e.g. Al with COOH or OH groups, Ti with COOCH$_3$, OH or CN groups….) (Herdt and Czanderna, 1995, Jung and Czanderna, 1994, Jung et al., 1996, Fisher et al., 2000, Fisher et al., 2002, Konstadinidis et al., 1995), a chemical reaction occurs forming a molecular overlayer on top of the monolayer. This overlayer made of organometallic complexes or metal oxides may perturb the electronic coupling between the metal and the molecule, leading, for instance, to partial or total Fermi-level pinning at the interface (Lenfant et al., 2006). In some cases, if the metal chemically reacts with the end-group of the molecule (e.g. Au on thiol-terminated molecules), this overlayer may further prevent the diffusion of metal atoms into the organic monolayer (Aswal et al., 2005). The metal/organic interface interactions (e.g. interface dipole, charge transfer,…) are very critical and they have strong impacts on the electrical properties of the molecular devices. Some reviews are given in Refs. (Cahen et al., 2005, Kahn et al., 2003). If the metal atoms are not too reactive (e.g. Al with CH$_3$ or OCH$_3$…) (Herdt and Czanderna, 1995, Jung and Czanderna, 1994, Jung et al., 1996, Fisher et al., 2000, Fisher et al., 2002, Konstadinidis et al., 1995), they can penetrate into the organic monolayer, diffusing to the bottom interface where they can eventually form an adlayer between this electrode and the monolayer (in addition to metallic filamentary short circuits). In a practical way for device application using organic monolayers, the metal evaporation is generally performed onto a cooled substrate (~100 K). It is also possible to intercalate blocking baffles on the direct path between the crucible and the sample, or/and to introduce a small residual pressure of inert gas in the vacuum chamber of the evaporator (Okazaki and Sambles, 2000, Metzger et al., 2001, Xu et al., 2001). These techniques allow reducing the energy of the metal atoms arriving on the monolayer surface, thus reducing the damages.

To avoid these problems, alternative and soft metal deposition techniques were developed. One called nanotransfer printing (nTP), has been described and demonstrated (Loo et al., 2003). Nanotransfer printing is based on soft lithographic techniques used to print patterns with nanometric resolution on solid substrates (Xia and Whitesides, 1998). The principle is briefly described as follows. Gold electrodes are deposited by evaporation onto an elastomeric stamp and then transferred by mechanical contact onto a thiol-functionalized SAM. Transfer of gold is based on the affinity of this metal for thiol function –SH forming a chemical bond Au–S. Loo et al. (Loo et al., 2003) have used the nTP technique to deposit gold electrodes on alkane dithiol molecules self-assembled on gold or GaAs substrates. Nanotransfer printing of gold electrodes was also deposited onto oxidized silicon surface covered by a monolayer of thiol-terminated alkylsilane molecules (Loo et al., 2002, Guerin et al.). Soft depositions of pre-formed metal electrodes, e.g. lift-off float-on (LOFO) (Vilan and Cahen, 2002), have also been developped. Recently, another solution has been proposed in which a thin conducting polymer layer has been intercalated as a buffer layer between the organic monolayer and the evaporated metal electrode (Akkerman et al., 2006). It was also reported to use metallic electrode made of a 2D network of carbone nanotubes (He et al., 2006). Finally, another solution to avoid problems with metal evaporation is to cover a metal wire (about 10 µm in diameter) with a SAM and then to bring this wire in contact with another wire (crossing each other) using the Laplace force (Kushmerick et al., 2002a, Kushmerick et al., 2002b). About $10^3$ molecules can be contacted by this way.

At the nanometer-scale, the top electrode can also be a STM tip. The properties of a very small number of molecules (few tens down to a single molecule) can be measured. If one assumes that an intimate contact is provided by the chemical grafting (in case of a SAM) at one end of the molecules on the bottom electrode, the drawback of these STM experiments is the fact that the electrical "contact" at the other end occurs through the air-gap between the SAM surface and the STM tip (or vacuum in case of an UHV-STM). This leads to a difficult estimate of the true conductance of the molecules, while possible through a careful data analysis and choice of experimental conditions (Bumm et al., 1999, Labonté et al., 2002). Recently, some groups have used a conducting-atomic force microscope (C-AFM) as the upper electrode (Wold and Frisbie, 2000, Wold and Frisbie, 2001, Wold et al., 2002). In that case, the metal-coated tip is gently brought into a mechanical contact with the monolayer surface (this is monitored by the feed-back loop of the AFM apparatus) while an external circuit is used to measure the current-voltage curves. The advantage over the STM is twofold, (i) tip-surface position control and current probing are physically separated (while the same current in the STM is used to control the tip position and to probe the electronic transport properties), (ii) under certain conditions, the molecules may be also chemically bounded to the C-AFM tip at the mechanical contact (Cui et al., 2001). The critical point of C-AFM experiments is certainly the very sensitive control of the tip load to avoid excessive pressure on the molecules (Son et al., 2001) (which may modify the molecule conformation and thus its electronic transport properties, or even can pierce the monolayer). On the other hand, the capability to apply a controlled mechanical pressure on a molecule to change its conformation is a powerful tool to study the relationship between conformation and electronic transport (Moresco et al., 2001). A significant improvement has been demonstrated by Xu and Tao (Xu and Tao, 2003) to measure the conductance of a single molecule by repeatedly forming few thousands of Au-molecule-Au junctions. This technique is a STM-based break junction, in which molecular junctions are repeatedly formed by moving back and forth the STM tip into and out of contact with a gold surface in a solution containing the molecules of interest. A few molecules, bearing two chemical groups at their ends, can bridge the nano-gap formed when moving back the tip from the surface. Due to the large number of measurements, this technique provides statistical analysis of the conductance data. This technique has been recently used to obtain new insights on the electronic transport through molecular junctions, e.g. on the analysis of the variability of the conductance (Ulrich et al., 2006, Venkataraman et al., 2006b), on the role of the chemical link between the molecule and the metal electrode (Chen et al., 2006, Venkataraman et al., 2006b) (for instance, it has been shown that the amine group gives a better defined conductance than thiol (Venkataraman et al., 2006b)), on the influence of the atomic configuration of the chemical link (Li et al., 2006). Changes in the electrical conductance of a single molecule as function of a chemical substitution (Venkataraman et al., 2007) and a conformational change were also evidenced (Venkataraman et al., 2006a).

The second type of MmM junctions uses a "planar" configuration (two electrodes on the same surface). The advantage over a vertical structure is the possibility to easily add a third gate electrode (3-terminal device) using a bottom gate transistor configuration. The difficulties are (i) to make these electrodes with a nanometer-scale separation; (ii) to deposit molecules into these nano-gaps. Alternatively, if the monolayer is deposited first onto a suitable substrate, it would be very hard to pattern, with a nanometer-scale resolution, the electrodes on top of it. The monolayers have to withstand, without damage, a complete electron-beam patterning process for instance. This has been proved possible for SAMs of alkyl chains (Collet et al., 2000, Collet and Vuillaume, 1998) and alkyl chain functionalized by $\pi$-conjugated oligomers (Mottaghi et al., 2007) used in nano-scale (15 – 100 nm) devices. However, recently developed soft-lithographies (micro-imprint contact…) can be used to pattern organic monolayers or to pattern electrodes on these monolayers (Xia and Whitesides, 1998). Nowadays, 30 nm width nano-gaps are routinely fabricated by e-beam lithography and 5 nm width nano-gaps are attainable with a lower yield (a few tens %) (Cholet et al., 1999, Bezryadin and Dekker, 1997, Guillorn et al., 2000). However, these widths are still too large compared to the typical molecule length of 1-3 nm.

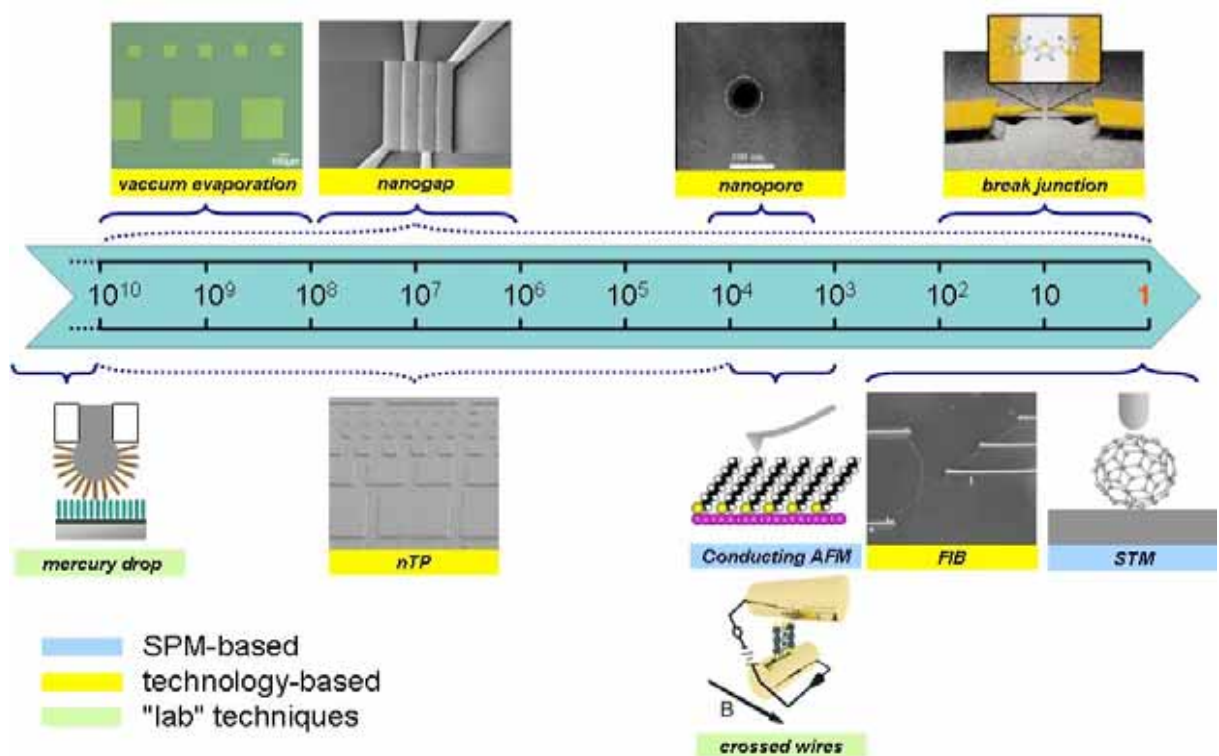

Fig. 2: A schematic overview of the different test-beds used to electrically contact organic molecules. The scale gives the approximate number of molecules contacted from monolayer (left) to single molecule (right). The techniques are (from left to right, upper part of the figure) : micrometer-scale metal evaporation, nano-gap patterned by e-beam lithography, nanopores, break-junction, and (from left to right, lower part of the figure) : mercury drop, nano-transfer printing, conducting AFM, crossed

wires, metal deposition by FIB, STM (courtesy of S. Lenfant, IEMN-CNRS).

The smallest nanogaps ever fabricated have a width of about 1 nm. A metal nanowire is e-beam fabricated and a small gap is created by electromigration when a sufficiently high current density is passing through the nanowire (Park et al., 1999). These gold nanogaps were then filled with few molecules (bearing a thiol group at each ends) and Coulomb blockade and Kondo effects were observed in these molecular devices (Liang et al., 2002, Park et al., 2002). A second approach is to start by making two electrodes spaced by about 50-60 nm, then to gradually fill the gap by electrodeposition until a gap of few nanometers has been reached (Boussaad and Tao, 2002, Li et al., 2000, Kervennic et al., 2002). Recently, carbone nanotubes (CNT) have been used as electrodes separated by a nano-gap (<10 nm) (Guo et al., 2006). The nano-gap is obtain by a precise oxidation cutting of the CNT, and the two facing CNT ends which are now terminated by carboxylic acids, are covalently bridged by molecules of adapted length derivatized with amine groups at the two ends. It is also possible to functionalize the molecule backbone for further chemical reactions allowing the electrical detection of molecular and biological reactions at the molecule-scale (Guo et al., 2006, Guo et al., 2007). Another approach is to use a breaking junction, bridged by few dithiol-terminated molecules. Reed and coworkers (Reed et al., 1997) and Kergueris and coworkers (Kergueris et al., 1999) have used these breaking junctions to fabricate and to study some MmM junctions based on dithiolbenzene and bisthiolterthiophene, respectively, and this technique was further used with others short oligomers (Reichert et al., 2002, Weber et al., 2002). However, these MmM breaking junctions are not stable over a very long period of time (no more than 20-30 min) while the vertical MmM junctions and the "planar" ones based on nanofabricated nano-gaps are stable over months. Weber et al. reported some improvements allowing stable MmM breaking junction measurements at low temperature (Elbing et al., 2005, Reichert et al., 2003). Finally, we mention that Au nanoparticles (NP) can be used to connect a few molecules, these NP (tens of nm in diameter) being themselves deposited between electrodes or contacted with a STM (Dadosh et al., 2005, Cui et al., 2001, Long et al., 2005). Microspheres metallized by Ni/Au can also be magnetically trapped between micro-lithographically patterned electrodes covered by a monolayer of molecules forming two molecular junctions in series (Long et al., 2005). These approaches allow measuring a small number of molecules and avoid the difficult fabrication of few nm size gaps. A very recent review on how to electrically connect molecules and organic monolayers is given by Haick and Cahen (Haick and Cahen, 2008).

To conclude this section, many technological solutions are available to measure the electronic transport properties of molecular monolayers with lateral extension from few molecules to ~$10^{10}$ (Fig. 2). A comparison between electrical measurements at the molecular-scale and those on macroscopic devices will be helpful to understand the effect of intermolecular interactions on the transport properties. As a result of these various approaches for making the organic monolayers and the MmM junctions, the nature of the interfaces, and thus the electronic coupling between the molecules and the electrodes are largely depending on the experimental conditions and protocols. This feature requires a multi test-bed approach to assess the intrinsic properties of the molecular devices and not of the contacts (Szuchmacher Blum et al., 2005). In the following sections, we illustrate and discuss the effects of this molecule/electrode coupling on the electronic transport properties of some molecular devices.

## 3. MOLECULAR TUNNELING BARRIER

It has long been recognized that a monolayer of alkyl chains sandwiched between two metal electrodes acts as a tunneling barrier. Mann and Kuhn (Mann and Kuhn, 1971), Polymeropoulos and Sagiv (Polymeropoulos, 1977, Polymeropoulos and Sagiv, 1978) have demonstrated that the current through LB monolayers of alkyl chains follows the usual distance-dependant exponential law, $I=I_0 exp(-\beta d)$, where d is the monolayer thickness and $\beta$ is the distance decay rate. They have found $\beta \sim 1.5$ Å$^{-1}$. More recently, we found (Lenfant, 2001) $\beta \sim 0.7$-$0.8$ Å$^{-1}$ for n$^+$-Si/native SiO$_2$/SAM of alkyl-1-enyl trichlorosilane/metal (Au or Al) junctions and Whitesides's group (Holmlin et al., 2001) found $\beta \sim 0.9$ Å$^{-1}$ for Hg/SAM of alkylthiol/Ag junctions. All these experiments were done with macroscopic-size electrodes. Data taken for alkanethiols in a nanopore junction gave $\sim 0.8$ Å$^{-1}$ (Wang et al., 2003). Recently, C-AFM experiments were also done addressing the properties of a small number of molecules. Again, a tunneling law was observed with $\beta \sim 0.9$-$1.4$ Å$^{-1}$ for Au/SAM of alkylthiols/Au-covered AFM tip junctions (Wold and Frisbie, 2000, Wold and Frisbie, 2001, Sakaguchi et al., 2001, Engelkes et al., 2004). A quite smaller value ($\beta \sim 0.5$ Å$^{-1}$) was reported for Au/SAM of alkyldithiol/Au-covered AFM tip junctions (Cui et al., 2002), but another work reported no significant variation of $\beta$ between alkanethiols and alkanedithiols, but only a contact resistance 1 or 2 decades lower for the alkanedithiols. A more complete review of these data and others is given in Ref. (Salomon et al., 2003). The $\beta$ value is related to the tunneling barrier height ($\Delta$) at the molecule/electrode interface and to the effective mass (m*) of carriers in the monolayer, $\beta=\alpha(m^*/m_0)^{1/2}\Delta^{1/2}$, with m$_0$ the rest mass of the electron and $\alpha = 4\pi(2m_0 e)^{1/2}/h = 10.25$ eV$^{-1/2}$ nm$^{-1}$ (e is the electron charge and h the Planck constant). The tunneling barrier height may be measured independently by internal photoemission experiment (IPE) (Powell, 1970) where carriers in one of the electrodes are photoexcited over the tunneling barrier and collected at the other electrode (under a small applied dc bias). Threshold energy of the exciting photons allows the measurement of $\Delta$. We have found an electron tunneling barrier of about 4.3-4.5 eV at the silicon/native SiO$_2$/SAM and aluminum/SAM interfaces in the case of densely packed, well-ordered, SAMs of alkyl chains (Boulas et al., 1996), a larger value than $\sim 1.4$ to 3 eV found in other experiments on LB monolayers and alkylthiol SAM on Au (Mann and Kuhn, 1971, Polymeropoulos and Sagiv, 1978, Holmlin et al., 2001, Wang et al., 2003). This high value ($\sim 4.5$ eV) is in agreement with theoretical calculations (Vuillaume et al., 1998). For the same alkyl chains directly chemisorbed on Si (no native oxide), lower values have been reported from a combination of electrical ($\sim 1$-$1.5$ eV) and UPS/IPES (2.5-3.5 eV) experiments (Salomon et al., 2005, Salomon et al., 2007). The discrepancy between electrical and spectroscopy data is due to the fact that charge carrier transport is dominated by the presence of interface states localized between the molecular HOMO (highest occupied molecular orbital) and LUMO (lowest unoccupied molecular orbital) and the Si band edges (Salomon et al., 2007).

These puzzling data may be rationalized if we consider the nature of the molecule/electrode coupling. Figures 3 shows some of these data in a $\beta$-$\Delta$ plot. The smallest β and Δ values are obtained for a good or "intimate" coupling at both the two electrodes. This is the case for SAM of alkyldithiols chemisorbed at the two electrodes (Cui et al., 2002), (Holmlin et al., 2001) and for SAM chemisorbed at one end and contacted at the other one by an evaporated metal (Lenfant, 2001). This is also the case for alkyl chains directly attached to Si without native oxide between the substrate and the molecules (Salomon et al., 2005, Salomon et al., 2007). The largest values are obtained when at least one coupling is weak, as it is the case for physisorbed LB monolayers (Mann and Kuhn, 1971, Polymeropoulos, 1977, Polymeropoulos and Sagiv, 1978) and SAM mechanically contacted by C-AFM tip (Wold and Frisbie, 2000, Wold and Frisbie, 2001) or chemisorbed on the native oxide of the Si substrate (Boulas et al., 1996, Vuillaume et al., 1998). In this latter case, the top metal electrode (Al or Au) was also weakly coupled with the $CH_3$-terminated molecules. The tunnel barrier height is lowered (2.2-2.5 eV) (Aswal et al., 2005) if Au is used as the top electrode on thiol-terminated SAM of alkyl chains still grafted on naturally oxidized Si, probably due to a better molecule/metal coupling through the S-Au chemical link. This feature reveals that the nature of the molecule/electrode coupling strongly changes the electronic properties of the molecules. The HOMO-LUMO gap of the molecule, and therefore the tunnel barrier height, may be reduced by several eV for a chemisorbed molecule on metal compared to the gas phase molecule (Vondrak et al., 1999). Charge transfer and interface dipole also move the position of the molecular orbitals with respect to Fermi energy of the electrodes. A review on these phenomena is given in Refs. (Cahen et al., 2005, Kahn et al., 2003). The molecule/electrode contact is a key parameter in the overall transport properties of the MmM junctions. It was demonstrated that the conductance of a MmM junction is increased when the molecule is chemisorbed at its two ends (via a thiol link on gold for instance) compared to the situation when only one end is chemically connected to one electrode. An increase by a factor $10^3$ was observed for a monolayer of octadecanedithiol molecules as compared to a monolayer of octadecanethiol (Cui et al., 2001, Beebe et al., 2002). Another experimental evidence is given by a comparison of two systems (Hg-S-alkyl and Hg/alkyl) where the sulfur linked molecules showed a better electrical conductivity (Selzer et al., 2002a).

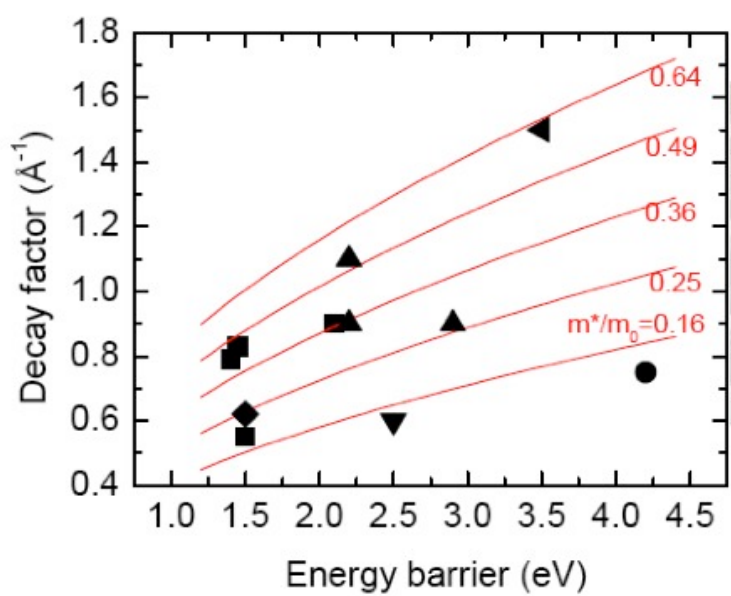

Fig. 3. Left: Tunnel decay factor – energy barrier plot for several molecular tunnel junctions: (■) metal-alkylthiol or dithiol-metal (Au or Hg) junctions (Holmlin et al., 2001, Wang et al., 2003, Cui et al., 2001), (▲) Au-alkylthiol or dithiol-Au C-AFM junctions (Beebe et al., 2002, Engelkes et al., 2004, Wold and Frisbie, 2001, Wold et al., 2002), (◄) LB monolayer (Mann and Kuhn, 1971), (♦) Si-alkyl-Hg junction (Salomon et al., 2005, Salomon et al., 2007), (●) Si-native $SiO_2$-alkylsilane-Al junction (Boulas et al., 1996, Vuillaume et al., 1998), (▼) Si-native $SiO_2$-mercaptopropyltrimethoxysilane-Au junction (Aswal et al., 2005). Lines are calculated according to the classical equation (see text) for different values of the effective mass.

Finally, these tunnel junctions are also good prototypal devices to study more detailed phenomena such as: electron – molecular vibration coupling using inelastic electron tunnel spectroscopy (IETS) (Kushmerick et al., 2004, Wang et al., 2004, Petit et al., 2005, Aswal et al., 2006, Beebe et al., 2007, Long et al., 2006), current-induced local heating in a molecular junction (Huang et al., 2006), dynamical charge fluctuations using noise measurements (Clement et al., 2007) and spin-polarized transport (Petta et al., 2004, Wang and Richter, 2006). Beyond the first results, more of such experiments are now required to achieve a good agreement between a variety of different results, as well as with theoretical predictions. These approaches open very interesting pathway toward a better understanding of electronic transport in molecular junctions.

4. MOLECULAR SEMICONDUCTING WIRE

Contrary to the case of fully saturated alkyl chains, short oligomers of π conjugated molecules are considered as the prototype of molecular semiconducting wires. At low bias, when the LUMO and HOMO of the molecules are not in resonance within the Fermi energy window opened between the two electrodes by the applied bias, the conduction is still dominated by tunneling. However, the decay factor β is lower than in the case of alkyl chains (see *supra*), typically β ~ 0.2 to 0.6 Å$^{-1}$. This is related to the lower HOMO-LUMO gap of the π-conjugated molecules (~ 2 – 4 eV, typically, against 8-9 eV for alkyl chains), and therefore to a lower energy barrier for charge injections. A detailed comparison of transport properties between saturated and π-conjugated molecules is given in ref. (Salomon et al., 2003) Bumm and coworkers (Bumm et al., 1996) have studied the conductivity of prototypes of molecular wires. A few molecules of di(phenylene-ethynylene)benzenethiolate were inserted in a SAM of dodecanethiols (which are insulating molecules), and the difference in conductivity was investigated using the tip of a STM. With a STM working at a constant current, the tip is retracted when passing over a more conducting molecule than the surrounding matrix of alkyl chains. Thus the apparent amplitude height in the STM image is directly related to the conducting behavior of these molecules. Patrone and coworkers (Patrone et al., 2002, Patrone et al., 2003b) have repeated these experiments for thiolterthiophene molecules, another prototype of molecular wires (Fig. 4). However, as explained *supra*, the drawback of these experiments is the fact that the electrical "contact" at the upper end of the molecules occurs through the air-gap between the SAM surface and the STM tip (or vacuum in case of an UHV-STM). This leads to a difficult estimation of the true conductance of the molecules. Reed et al. (Reed et al., 1997), Kergueris et al. (Kergueris et al., 1999), Weber et al. (Reichert et al., 2002, Weber et al., 2002, Elbing et al., 2005) have used breaking junctions to fabricate and to study some MmM junctions based on short conjugated oligomers. The current-voltage curves are strongly non-linear with steps (peaks in the first derivative) corresponding to resonant charge carrier transfer through the molecular orbitals (MO) of the molecules. The measured conductance corresponds to the conductance through the molecules and the conductance of the molecule/electrode contact. Thus, the influence of the chemical link between the molecules and the electrode is of a prime importance. A change from an asymmetric to a symmetric current-tension (with respect to the bias polarity) curve was observed when comparing MmM junctions of SAMs of monothiolate and dithiolate oligo(phenylene ethynylene) molecules (Kushmerick et al., 2002b). The current increases by about a factor 10 when a sulfur atom attaches the molecule to the gold electrode compared to a mechanical contact. Today, the thiol group is the most used link to gold. However, theoretical calculations have recently predicted that selenium (Se) and tellurium (Te) are better links than sulfur (S) for the electronic transport through MmM junctions based on phenyl-based molecular wires (Di Ventra and Lang, 2001, Yaliraki et al., 1999). This was recently demonstrated in a series of experiments using SAMs made of bisthiol- and biselenol-terthiophene molecules inserted in a dodecanethiol matrix (Patrone et al., 2002, Patrone et al., 2003a). Using both STM in ambient air and UHV-STM, the apparent height of the molecular wires above the dodecanethiol matrix (as in the Bumm et al. work quoted above (Bumm et al., 1996)) is used to compare the electron transfer through the terthiophene molecule linked to the gold surface by S or Se atoms. Whatever the experimental conditions (air or UHV, tip-substrate bias, tunnel current set-point), the Se-linked molecules always appear higher in the STM images than the ones with a S linker. This feature directly demonstrates that a Se atom provides a better electron coupling between the gold electrode and the molecular wire than a S atom does (at least for the terthiophene molecule used in these experiments). From UPS experiments, this was attributed to a reduction of the energy offset between the highest occupied molecular orbital (HOMO) of the molecules (these molecules are mainly a better hole transport material than an electron transport material) and the Fermi energy of the gold electrode (Patrone et al., 2002, Patrone et al., 2003b). This offset reduction is in agreement with theory (Di Ventra and Lang, 2001, Yaliraki et al., 1999). Similarly, comparing the electron transport through SAMs of alkylthiols and alkyl-isonitriles (C-AFM measurements), it was established that the contact resistance for the Au/CN link is about 10% lower than for the Au/S interface (Beebe et al., 2002). Further experiments have shown that : i) amine group ($NH_2$) give better controlled conductance variability than thiol (SH) and isonitrile (CN) (Venkataraman et al., 2006b) and ii) the interface contact resistance is lower for amine than for thiol (Chen et al., 2006). Further experiments are now required to deeply investigate all possible anchoring atom/electrode couples (S, Se, Te, CN, COOH etc…, on one side and Au, Ag Pt, Pd, for instance, on the other side) and to determine to which extent the conclusions drawn for a peculiar molecule are valid for any other ones. With all these data on hands, one would optimize the design of future devices for molecular electronics. Electron-molecular vibronic coupling in short semiconducting oligomers has also been recently studied by IETS (Kushmerick et al., 2004, Long et al., 2006) as for alkane molecules, as well as thermoelectricity in these molecular junctions (Reddy et al., 2007). In this latter case, the Seebeck coefficient of the single molecules has been determined, as well as a clear evidence of hole transport through the junctions. This result allows beginning to explore thermoelectric energy conversion at the molecular-scale.

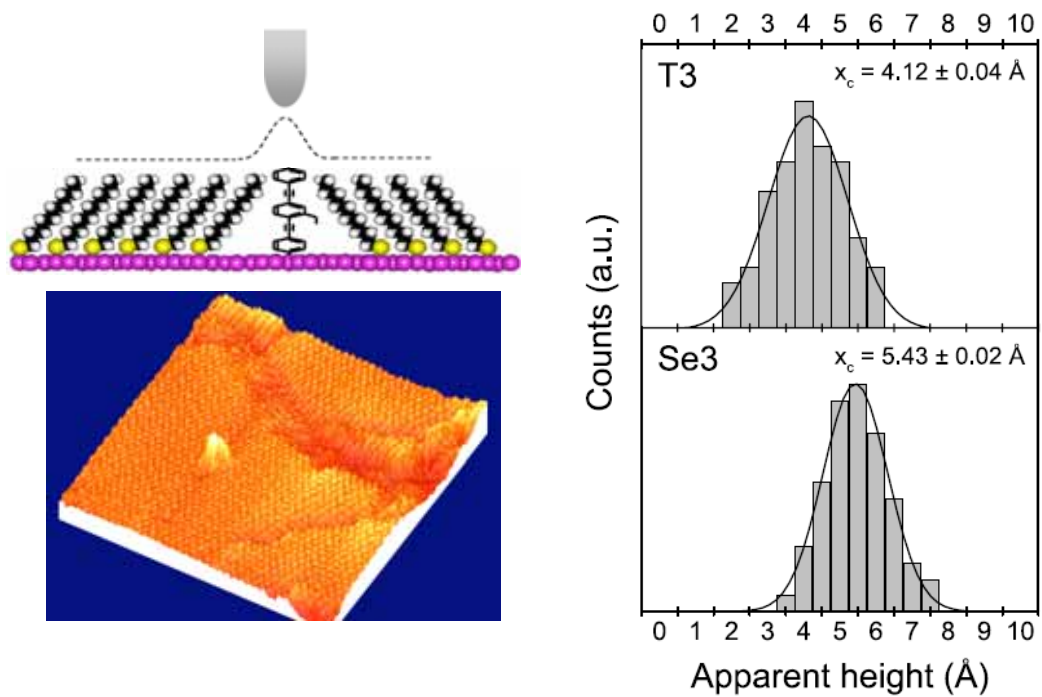

Fig. 4. Top-left, schematic view of a mixed monolayer where a few "conducting" molecules (dithiol-terthiophene) are intercalated into "insulating" ones (alkanethiol) used for STM measurements. Bottom-left, STM image (28 nm x 28 nm). The bump in the image is due to a higher current when the tip is passing over the more conducting terthiophene molecules. The background corresponds to the tunneling current through the alkanethiols (Patrone et al., 2002, Patrone et al., 2003b). Right, comparison of the apparent height (which is related to the molecular conductance) measured on the STM images for the S- and Se-linked terthiophene molecules – T3 and Se3, respectively (histogram taken from many measurements). Copyright (2002) with permission from Elsevier.

5. MOLECULAR RECTIFYING DIODE

A basic molecular device is the electrical current rectifier based on suitably engineered molecules. This molecular diode is the organic counterpart of the semiconductor p-n junction. At the origin of this idea, Aviram and Ratner (AR) proposed in 1974 to use D-σ-A molecules where D and A are respectively electron donor and acceptor, and σ is a covalent "sigma" bridge (Aviram and Ratner, 1974). Several molecular rectifying diodes were synthesized based on this AR paradigm, with donor and acceptor moieties linked by a short σ or even π bridge (Metzger, 1999, Metzger and Panetta, 1991, Ashwell et al., 1990, Martin et al., 1993, Metzger et al., 1997, Metzger et al., 2001, Vuillaume et al., 1999, Xu et al., 2001). This D-b-A (b=bridge) group is also ω-substituted by an alkyl chain to allow a monolayer formation by the Langmuir-Blodgett (LB) method and this LB monolayer is then sandwiched in a metal/monolayer/metal junction. The first experimental results were obtained with the hexadecylquinolinium tricyanoquinodimethanide molecule ($C_{16}H_{33}$-Q-3CNQ for short) – Fig. 5 (Metzger et al., 1997, Vuillaume et al., 1999, Metzger et al., 2001, Xu et al., 2001, Martin et al., 1993, Ashwell et al., 1990). However, the chemical synthesis of this molecule was not obvious with several routes leading to erratic and unreliable results. A more reliable synthesis was reported with a yield of 59% (Metzger et al., 1997). More recently, other D-b-A molecules have been synthesized and tested (Metzger et al., 2003, Baldwin et al., 2002) showing rectification with a ratio up to ~$2x10^4$. We can also mention some other approaches using D-A diblock co-oligomers (Ng et al., 2002) or CNT asymmetrically functionalized by D and A moieties at their ends (Wei et al., 2006) with a rectification ratio of ~$10^3$ in this latter case. Even if these results represent an important progress to achieve molecular electronics, the physical mechanism responsible for the rectification is not clear. One critical issue is to know if the AR model can be applied to $C_{16}H_{33}$-Q-3CNQ because it is a D-$\pi$-A molecule (Metzger et al., 1997), and due to the $\pi$ bridge, the HOMO and LUMO may be more delocalized than expected in the AR model. On the theoretical side, these molecular diodes are complex systems, characterized by large and inhomogeneous electric fields, which result from the molecular dipoles in the monolayer, the applied bias and the screening induced by the molecules themselves and the metallic electrodes. A theoretical treatment of these effects requires a self-consistent resolution of the quantum mechanical problem, including the effect of the applied bias on the electronic structure. Combining ab initio and semi-empirical calculations, it was shown (Krzeminski et al., 2001) that the direction of easy current flow (rectification current) depends not only on the placement of the HOMO and LUMO relative to the Fermi levels of the metal electrodes before bias is applied, but also on the shift induced by the applied bias: this situation is more complex than the AR mechanism, and can provide a rectification current in an opposite direction. The electrical rectification results from the asymmetric profile of the electrostatic potential across the system (Krzeminski et al., 2001, Stokbro et al., 2003). In other words, this means that the molecule is more strongly coupled with one electrode than with the other one (more closer to one of the electrodes due to the presence of the alkyl chain). The alkyl tail in the $C_{16}H_{33}$-Q-3CNQ molecule plays an important role in this asymmetry, and it was predicted (Krzeminski et al., 2001) a symmetric current-voltage curve in the case of molecules without the alkyl chain. This asymmetry effect was further theoretically studied more extensively (Kornilovitch et al., 2002, Taylor et al., 2002). Generally speaking, any asymmetric coupling of the molecules with the electrodes or any asymmetry in the molecule will result in a rectification effect (Datta et al., 1997, Elbing et al., 2005) – Fig. 5. This emphasizes the importance of the electrostatic potential profile in a molecular system and suggests that this profile can be chemically engineered to build new devices. For instance, based on these considerations, we have recently reported an experimental demonstration of a simplified and more robust synthesis of a molecular rectifier with only one donor group and an alkyl spacer chain (Lenfant et al., 2003, Lenfant et al., 2006). We have used a sequential self-assembly process (chemisorption directly from solution) on silicon substrates. We have analyzed the properties of these molecular devices as a function of the alkyl chain length and for ten different donor groups. We have obtained rectification ratios up to 37 (Fig. 5). We have shown that rectification occurs from resonance through the HOMO of the $\pi$-group in good agreement with our calculations and internal photoemission spectroscopy. However, improvements are still required to suppress Fermi-level pinning at the molecule/metal interface (Lenfant et al., 2006) and to allow a clear design and tuning of the electrical behavior of the molecular diode through the right choice of the chemical nature of the molecule. This approach will allows us to fabricate molecular rectifying diodes compatible with silicon nanotechnologies for future hybrid circuitries. Finally, more efforts have been also put forward to design and synthesis new D-b-A molecules not affected by the presence of an asymmetric alkyl chain (see Fig. 5 for one example).(Baldwin et al., 2002, Honciuc et al., 2007, Metzger et al., 2003)

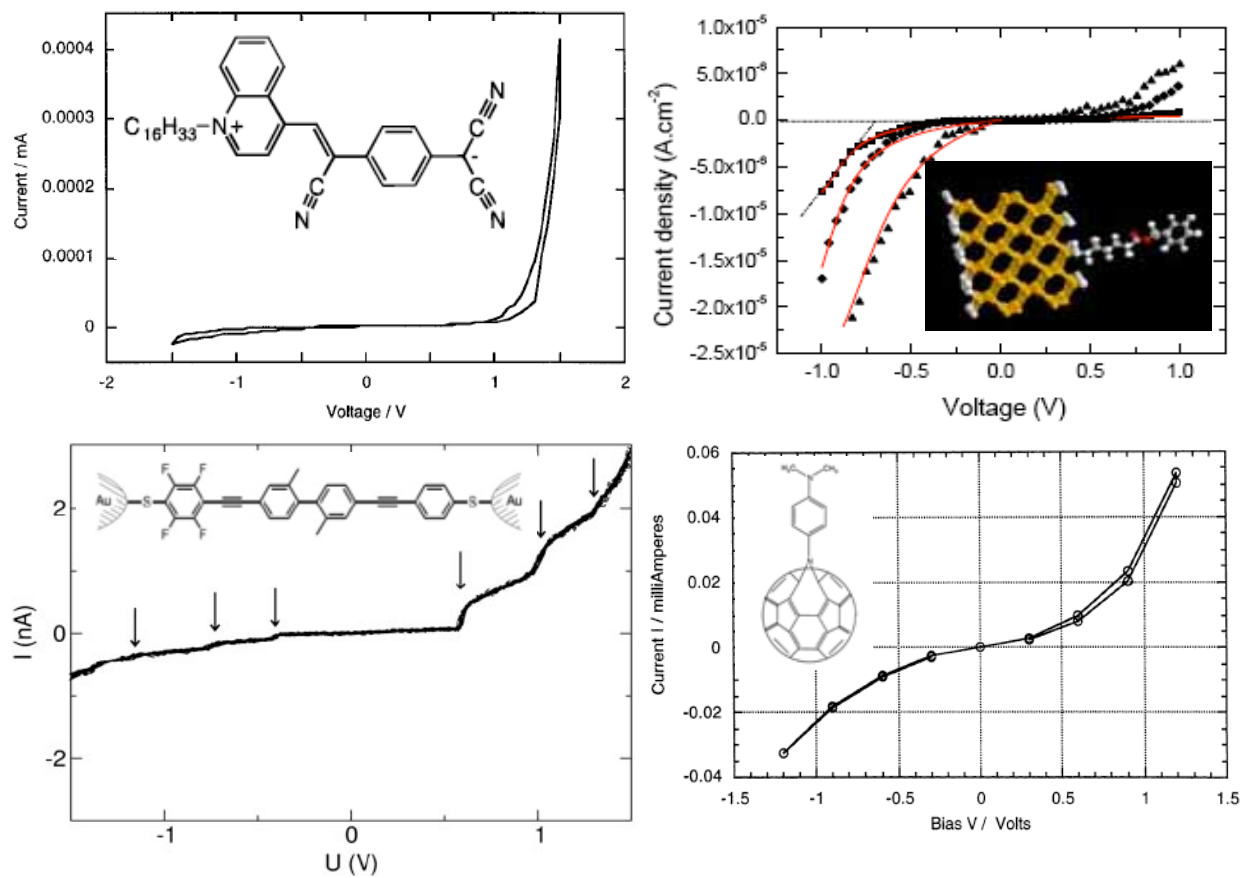

Fig. 5. Typical current-voltage characteristics of some molecular rectifying diodes. From top to bottom: LB monolayer of D-π-A molecules (tricyanoquinodimathanide) between metal electrodes, from (Metzger et al., 1997, Vuillaume et al., 1999), σ-π molecule grafted on Si (σ is alkyl chain and π groups are thiophene (▲ and ♦) and phenyl (■)) , from (Lenfant et al., 2006, Lenfant et al., 2003), D-A molecule inserted in a break-junction (at 30K in this latter case), from (Elbing et al., 2005) (copyright 2005, National Academy of Science, USA), and D-b-A (dimethylanilinoazafullerene) LB monolayers, from (Metzger et al., 2003).

## 6. MOLECULAR SWITCHES AND MEMORIES

Molecular switches and memories were also suggested at the early stage of the molecular electronics history (Aviram et al., 1988, Aviram et al., 1989, Aviram, 1988). We generally distinguish three approaches called "conformational memory", "charge-based memory" and 'RTD-based memory" (RTD is resonant tunneling diode). The first one relies on the idea to store a data bit on two bistable conformers of a molecule; the second on different redox states and the third on a negative differential resistance (NDR) due to resonant tunneling through molecular orbitals.

6.1.    Conformational memory

One of the most interesting possibilities for molecular electronics is to take advantage of the soft nature of organic molecules. Upon a given excitation, molecules can undergo conformational changes. If two different conformations are associated with two different conductivity levels of the molecule, this effect can be used to make molecular switches and memories. Such an effect is expected in π-conjugated oligomers used as molecular wires, if one of the monomer is twisted away from a planar conformation of the molecule (Venkataraman et al., 2006a). Twisting one monomer breaks the conjugation along the backbone, thus reducing the charge transfer efficiency along the molecule. This has been experimentally observed for a small molecular wire where the central unit was substituted with redox moieties. With the nanopore configuration to fabricate the MmM junction, Chen and coworkers (Chen et al., 1999, Chen et al., 2000) have observed that molecules with a nitroamine redox center (2'-amino-4,4'-di(ethynylphenyl)-5'-nitro-1-benzenethiol) exhibit a negative differential resistance behavior. In other words, they have observed that for a certain voltage range (typically between 1.5 and 2.2 V) applied on the MmM junction, the conductivity of the junction increased by a factor $10^3$ (At 60K, while the on/off ratio dropped to 1 at about 140 K. Other molecules with some changes of the redox moieties have exhibited on/off ratio of about 1.5 at RT (Chen et al., 2000)). They have also reported the feasibility of molecular random access memory cell using these molecules (Reed et al., 2001). The switching behavior of these compounds inserted in an alkanethiol SAM was also observed by STM (Donhauser et al., 2001). To separate the intrinsic behavior of the molecules from the molecule/metal interface, the same types of molecules have been measured on various test-beds (Fig. 6) (Szuchmacher Blum et al., 2005). These experiments demonstrated a clear bias-induced switching, while with a large statistical variability. However, it is not firmly established that this switching behavior is solely due to the molecules. Recently, the Lindsey's group showed that another possible mechanism is a random and temporary break in the chemical link between the molecule and the gold surface (Ramachandran et al., 2003) and this point is still a subject of debate.

   Catenane and rotaxane are a class of molecules synthesized to exhibit a bistable bahavior. In brief, these molecules are made of two parts, one allowed to move around or along the other one (e.g. a ring around a rod, two interlocked rings). These molecules adopt two different conformations depending on their redox states, changing the redox state triggers the displacement of the mobile part of the structure to minimize the total energy. This kind of molecules was used to build molecular memories. A MmM junction using a LB monolayer of these molecules mixed with phospholipid acid showed a clear electrical bistable behavior at room temperature (Collier et al., 2001, Collier et al., 2000, Pease et al., 2001). A voltage pulse of about 1.5 - 2 V was used the switch the device from the "off" state to its "on" state.

The state was read at a low bias (typically 0.1-0.2 V). The on/off ratio was about a few tens. A pulse in reverse bias (-1.5 to -2 V) returned the device to the "off" state. Using these molecular devices, Chen and coworkers (Chen et al., 2003a, Chen et al., 2003b) have demonstrated a 64 bits non-volatile molecular memory cross-bar with an integration density of 6.4 Gbit/cm$^2$ (a factor ~10 larger than the state-of-the-art today's silicon memory chip). The fabrication yield of the 64 bits memory is about 85%, the data retention is about 24 h and about 50-100 write/erase cycles are possible before the collapse of the on/off ratio to 1. Recently a 160 kbit based on the same class of molecules has been reported, patterned at a 33 nm pitch ($10^{11}$ bits/cm$^2$) (Green et al., 2007). About 25% of the tested memory points passed an on/off ratio larger than 1.5 with an average retention time of ~ 1h. However, it has also been observed that similar electrical switching behaviors can be obtained without such a class of bistable molecules (i.e. using simple alkyl chains instead of the rotaxanes) (Stewart et al., 2004). The switching behavior is likely due to the formation and breaking of metallic micro-filaments introduced though the monolayer during the top metal evaporation. The presence of such filaments is not systematic (see discussion *supra*), however caution has to be taken before to definitively ascribe the memory effect as entirely due to the presence of the molecules. While having rather poor performances at the moment, these demonstrations allow us to envision the coming era of hybrid-electronics, where molecular cross-bar memories like these ones, will be addressed by multiplexer/demultiplexer and so one fabricated with standard semiconductor CMOS technologies (Chen et al., 2003a). The advantage of such molecular cross-bar memories are i) a low cost, ii) a very high integration density, iii) a defect-tolerant architecture, iv) an easy post-processing onto a CMOS circuitry and v) a low power consumption. For instance, it has been measured that an energy of ~50 zJ (or ~ 0.3 eV) is sufficient to rotate the dibutyl-phenyl side group of a single porphyrin molecule (Loppacher et al., 2003). This is ~$10^4$ lower than the energy required to switch a state-of-the art MOSFET, and near the kTLn2 (2.8 zJ at 300K, or 0.017 eV) thermodynamic limit.

6.2. Charge-based memory

The redox-active molecules, such as mettalocene, porphyrin and triple-decker sandwich coordination compounds attached on a silicon substrate have been found to act as charge storage molecular devices (Li et al., 2002, Roth et al., 2002, Roth et al., 2003, Liu et al., 2003). The molecular memory works on the principle of charging and discharging of the molecules into different chemically reduced or oxidized (redox) states. It has been demonstrated that porphyrins (i) offer the possibility of multibit storage at a relatively low potentials (below ~ 1.6 V), (ii) can undergo trillions of write/read/erase cycles, (iii) exhibit charge retention times that are long enough (minutes) compared with those of semiconductor DRAM (tens of ms) and (iv) are extremely stable under harsh conditions (400°C – 30 min) and therefore meet the processing and operating conditions required for use in hybrid molecule/silicon devices (Liu et al., 2003). Moreover, the same principle works with semiconducting nanowires dressed with redox molecules in a transistor configuration (Duan et al., 2002, Li et al., 2004a, Li et al., 2004b). Optoelectronic memories have also been demonstrated with polymer-functionalized CNT transistors (Borghetti et al., 2006, Star et al., 2004). However, in all cases, further investigations on the search of other molecules and, understanding the factors that control parameters such as, charge transfer rate, which limit write/read times, and charge retention times, which determines refresh rates, are needed.

6.3.  RTD-based memory

Memory can also be implemented from RTD devices following cell architecture already used for semiconductor devices. Memory cell based on RTD can be set up with 2 RTD and 2 transistors in a cross-bar architecture (Van Der Wagt et al., 1998) . The advantages compared to "resistive" and "capacitive" molecular memories are fast switching times and possible long retention times. RTD devices are characterized by a NDR behavior in their current-voltage curves. Many papers reported NDR behavior through molecular junctions (Chen et al., 2000, Chen et al., 1999, Li et al., 2003, Gorman et al., 2001, Amlani et al., 2002, Rawlett et al., 2002, Kratochvilova et al., 2002, Le et al., 2003) with peak-to-valley ration from about 1.5 to 5 (at room temperature). However a NDR may be also induced by other physical phenomena such as conformational changes already discussed *supra* or thiol-gold bond fluctuations (Ramachandran et al., 2003). The principle of a RTD molecular device is similar to that of his solid state counter-part (a potential well separated of the electrodes by two tunnel barriers). In the molecular analogue, the barriers should consist of aliphatic chains (of variable length) and the well should be made up of a short conjugated oligomer. Even if NDR behavior has been observed from STM results on single molecule attached to Si (Guisinger et al., 2004) and has been ascribed to resonance through the molecular orbitals in agreement with a theoretical result (Rakshit et al., 2004), this interpretation has been ruled out both experimentally (Pitters and Wolkow, 2006) and theoretically (Quek et al., 2007). In a detailed STM experiments, Pitters and Wolkow showed that the NDR behavior might be explained by random changes of the conformation of molecules on the surface (molecular rearrangement, desorption and/or decomposition) and not by a resonant tunneling through the molecule orbitals. These random phenomena are triggered by inelastic interactions between the molecular vibrations and electrons passing through the molecules. Resonant tunneling was also theoretically questionned by Quek et al. (Quek et al., 2007) using density functional theory and many-electron GW self-energy approach. They showed, for the specific molecule (e.g. cyclopentene on silicon) that the frontier energy levels do not move with the applied electric field, thus the molecular orbitals could not align with the silicon energy bands at certain applied bias. In conclusion, the exact origin of the molecular NDR behavior is still an open question, and therefore the RTD molecular device was not yet clearly demonstrated.

7.  MOLECULAR TRANSISTOR

A true transistor effect (i.e. the current through 2 terminals of the device controlled by the signal applied on a third terminal) embedded in a single three-terminal molecule (e.g. a star-shaped molecule) has not been yet demonstrated. Up to date, only hybrid-transistor devices have been studied. The typical configuration consists of a single molecule or an ensemble of molecules (monolayer) connected between two source and drain electrodes separated by a nanometer-scale gap, separated from an underneath gate electrode by a thin dielectric film – Fig. 7. At a single molecule level (single-molecule transistor), these devices have been used to study Coulomb blockade effects and Kondo effects at very low temperature. For instance,

Coulomb blockade (electron flowing one-by-one between source and drain through the molecule due to electron-electron Coulomb repulsion, the molecule acting as a quantum dot) was observed for molecules such as fullerene ($C_{60}$) and oligo-phenyl-vinylene (OPV) weakly coupled to the source-drain electrodes.(Park et al., 2000, Kubatkin et al., 2003) In this latter case, up to 8 successive charge states of the molecule have been observed. With organo-metallic molecules bearing a transition metal, such as Cobalt terpiridynil complex and divanadium complex, Kondo resonance (formation of a bound state between a local spin on the molecule, or an island, or a quantum dot, and the electrons in the electrodes leading to an increase of the conductance at low bias, around zero volt) has also been observed in addition to Coulomb blockade.(Park et al., 2002, Liang et al., 2002) Kondo resonance is observed when increasing the coupling between the molecule and the electrodes (for instance by changing the length of the insulating tethers between the metal ion and the electrodes). At a monolayer level, self-assembled monolayer field-effect transistors (SAMFET) have been demonstrated at room temperature.(Tulevski et al., 2004, Mottaghi et al., 2007) The transistor effect is observed only if the source and drain length is lower than about 50 nm, that is, more or less matching the size of domains with well organized molecules in the monolayer. This is mandatory to enhance $\pi$ stacking within the monolayer and to obtain a measurable drain current. SAM of tetracene,(Tulevski et al., 2004) terthiophene and quaterthiophene(Mottaghi et al., 2007) derivatives have been formed in this nano-gap. Under this condition, a field effect mobility of about $3.5 \times 10^{-3}$ $cm^2V^{-1}s^{-1}$ was measured for a SAMFET made with a quaterthiophene (4T) moiety linked to a short alkyl chain (octanoic acid) grafted on a thin aluminum oxide dielectric (Fig. 7). This value is on a par with those reported for organic transistor made of thicker films of evaporated 4T ($10^{-3}$ to $10^{-2}$ $cm^2V^{-1}s^{-1}$).(Mottaghi et al., 2007) The on/off ratio was about $2 \times 10^4$. For some devices, a clear saturation of the drain current vs. drain voltage curve has been observed, but usually, these output characteristics display a super linear behavior. This feature has been explained by a gate-induced lowering of the charge injection energy barrier at the source/organic channel interface.(Collet et al., 2000).

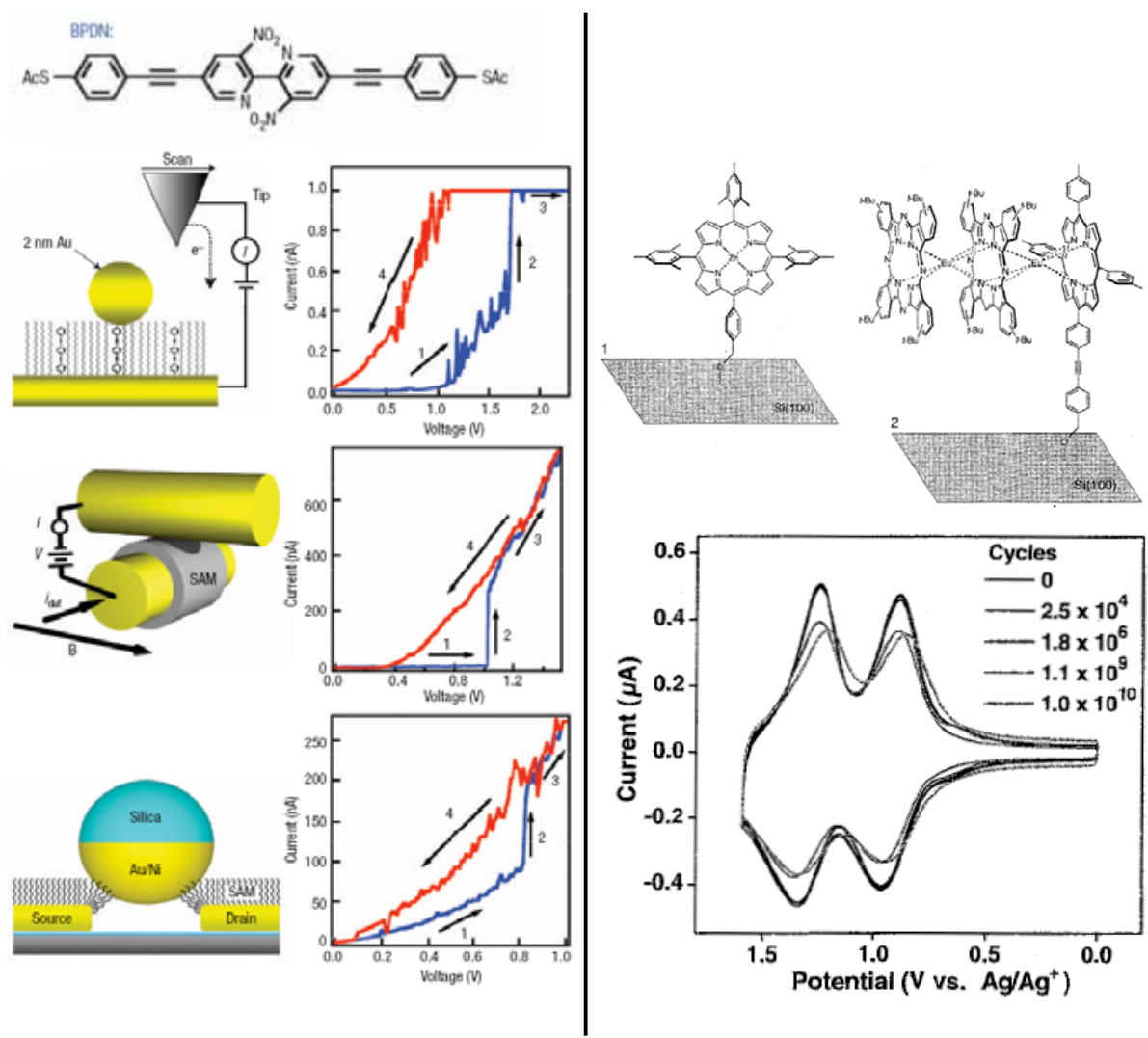

Fig. 6: Left: Current-voltage characteristics of bipyridyl-dinitro oligophenylene-ethynylene dithiol connected by Au electrodes using different test-beds (top to bottom): Au nanoparticle with STM, crossed-wires put in contact by the Lorentz force and Ni/Au metallized microsphere used as a magnetic bead junction. These experiments demonstrate a clear bias-induced switching behavior, while with a large variability. From (Szuchmacher Blum et al., 2005) – (reprinted by permission from Macmillan Publishers Ltd: Nature Materials, copyright 2005). Right: Typical redox molecules (porphyrin derivatives) attached to a silicon substrate used in a charge-based molecular memory device and its electrical response as a function of the number of write/erase cycles. This electrochemical response show 2 redox states that can be used to implement a multi-level memory, from (Liu et al., 2003) reprinted with permission from AAAS.

8. CONCLUSION.

We have described several functions and devices that have been studied at the molecular scale: tunnel barrier, molecular wire, rectifying and NDR diodes, bistable devices and memories. However, a better understanding and further improvements of their electronic properties are mandatory and need to be confirmed. These results suffer from a large dispersion and more efforts are now required to improve reproducibility and repeatability. For viable applications, more efforts are also mandatory to test the integration of molecular devices with silicon-CMOS electronics (hybrid molecular-CMOS nanoelectronics). Moreover most of these devices are 2-terminal, what's about a true/fully molecular 3-terminals device? We have also pointed out that the molecule-electrode coupling and conformation strongly modify the molecular-scale device properties. Molecular engineering (changing ligand atoms for example) may be used to improve or adjust the electrode-molecule coupling. Nevertheless, a better control of the interface (energetics and atomic conformation) is still compulsory. Beyond the study of single or isolated devices, more works towards molecular architectures and circuits are required. Up to now, mainly the « cross-bar » architecture has been studied. Is it sufficient? More new architectures must be explored (e.g. non von Neuman, neuronal, quantum computing…). Open questions concern the right approaches for inter-molecular device connections and nano-to-micro connections, the interface with the outer-world, hybridation with CMOS and 3D integration (Dehon et al., 2003, Tour et al.,

2002, Likharev and Strukov, 2005, Goldstein and Budiu, 2001). Beyond the CMOS probably bets on non-charge based devices. Molecular devices using other state variables (e.g. spin, molecule conformation,…) to code a logic state are still challenging and exciting objectives. Finally, other reviews, current status and challenges on charge transfer on the nanoscale can be found in (Adams et al., 2003, Nitzan, 2001, Tao, 2006, Nitzan and Ratner, 2003).

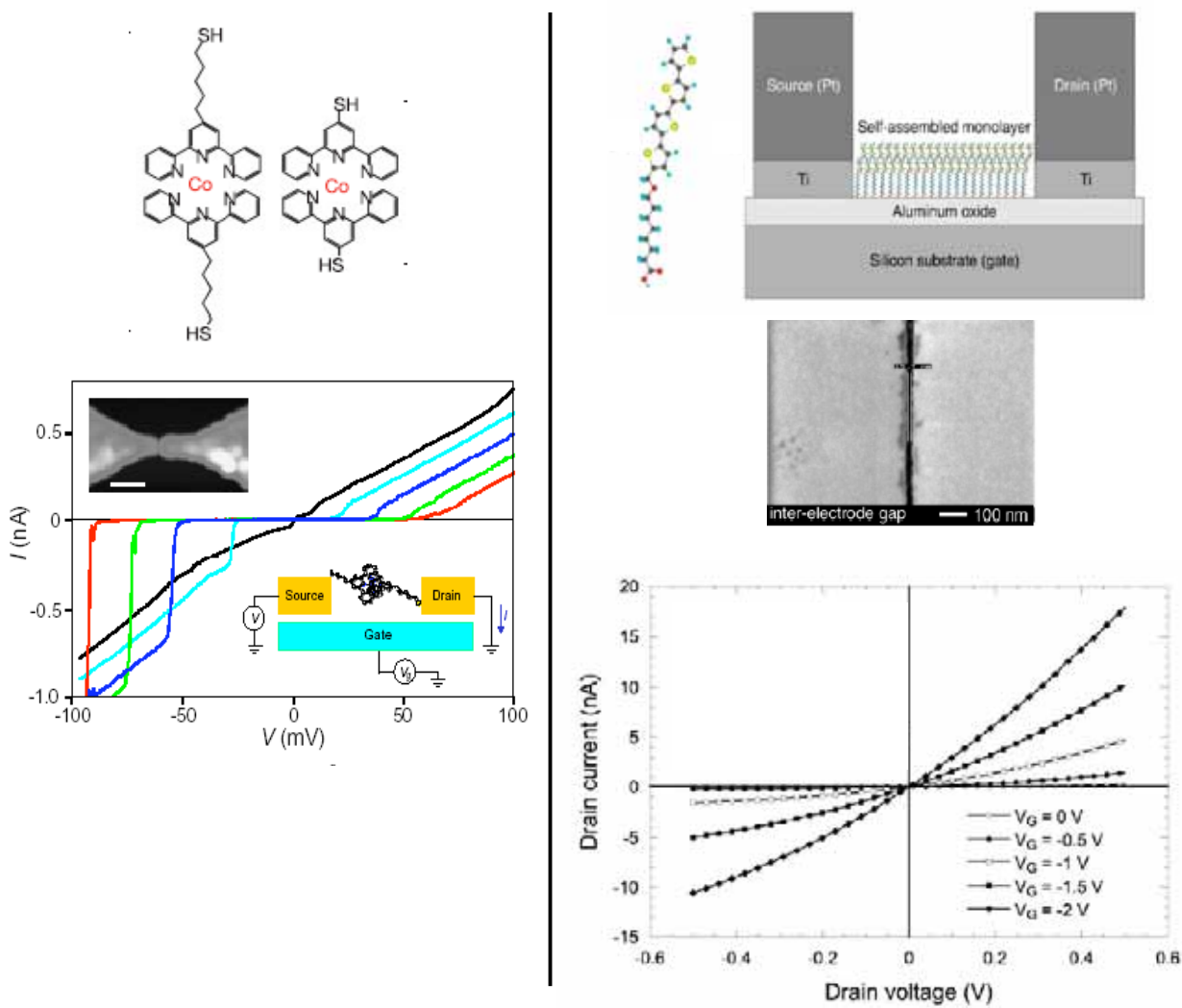

Fig. 7: Left: Structure of the Co-terpirydinyl complex molecules, AFM image of the source-drain nanogaps (~1-2nm) made by electromigration, and typical I-V with Coulomb blockade gaps measured at 100 mK for various gate voltage, and schematic diagram of the device; from (Park et al., 2002) – (reprinted by permission from Macmillan Publishers Ltd: Nature, copyright 2002). Right: Schematic diagram of the SAMFET and the 4T-octanoic acid molecule, SEM image of the 16 nm source-drain gap, and typical drain current-drain voltage curve for various gate voltage measured at 300 K, from (Mottaghi et al., 2007) - Copyright Wiley-VCH Verlag GmbH & Co. KGaA. Reproduced with permission.

ACKNOWLEDGMENTS

The works done at IEMN were financially supported by CNRS, ministry of research, ANR-PNANO, IRCICA, EU-FEDER Region Nord-Pas de Calais and IFCPAR. I thank all the colleagues in the "molecular nanostructures & devices" group at IEMN and many others outside our group for fruitful collaborations.